\documentclass[prd, amsfonts, twocolumn, nofootinbib, showpacs]{revtex4}
\usepackage{graphicx, epsfig}
\usepackage{color}
\usepackage{amsmath}
\usepackage{amssymb}
\usepackage{physics}
\usepackage{subfigure}
\newcommand{\be}{\begin{equation}}
\newcommand{\ee}{\end{equation}}
\newcommand{\bea}{\begin{eqnarray}}
\newcommand{\eea}{\end{eqnarray}}

\newcommand{\gapp}{\mathrel{\raise.3ex\hbox{$>$}\mkern-14mu
\lower0.6ex\hbox{$\sim$}}}
\newcommand{\lapp}{\mathrel{\raise.3ex\hbox{$<$}\mkern-14mu
\lower0.6ex\hbox{$\sim$}}}
\def\bbox{{\,\lower0.9pt\vbox{\hrule \hbox{\vrule height 0.2 cm
\hskip 0.2 cm \vrule  height 0.2 cm}\hrule}\,}}

\begin{document}
\title{Interaction of cosmological domain walls with large classical objects, like planets and satellites, and the 
flyby anomaly}
%\author{}
%\affiliation{ }
\author{De-Chang Dai$^{1,2}$\footnote{communicating author: De-Chang Dai,\\ email: diedachung@gmail.com\label{fnlabel}},  Djordje Minic$^3$, Dejan Stojkovic$^4$
}
\affiliation{$^1$ Center for Gravity and Cosmology, School of Physics Science and Technology, Yangzhou University, 180 Siwangting Road, Yangzhou City, Jiangsu Province, P.R. China 225002 }
\affiliation{ $^2$ CERCA/Department of Physics/ISO, Case Western Reserve University, Cleveland OH 44106-7079}
\affiliation{ $^3$ Department of Physics, Virginia Tech, Blacksburg, VA 24061, U.S.A. }
%\affiliation{ $^4$ }
\affiliation{ $^4$ HEPCOS, Department of Physics, SUNY at Buffalo, Buffalo, NY 14260-1500, U.S.A.}
 %%%%%%%%%%%%%%%%%%%%%%%%%%%%%%%%%%%%%%%%%%%%%%%%%%%%%%%

\begin{abstract}
\widetext
Cosmological domain walls can be formed as a result of symmetry breaking at any epoch during the evolution of our universe. We study their interaction with a classical macroscopic object, like Earth or a satellite in Earth's orbit. We set up an action that includes the interaction term between the massive classical object and the scalar field that the domain wall is made of. We use  numerical calculations to solve the coupled equations of motion which describe  the crossing between the domain wall and the classical object. Depending on the strength of the interaction, relative velocity and size, the object can be either stopped by the wall, or it can pass through it inducing deformations in the wall that cost energy. At the same time, the coupling to the scalar filed might change the object's  mass during the crossover. The fact that satellites in Earth's orbit (or planets in Sun's orbit) can change their mass and/or lose energy interacting with walls can be used as a new domain wall detection probe. For example, a typical velocity precision of a satellite is about $0.5$ mm/s, which directly puts an upper limit on its mass change to $\Delta M/M \lessapprox 5\times 10^{-17} $. Alternatively, a known satellite flyby anomaly can easily be explained as an interaction with a closed domain wall.  We also show that the presence of matter modifies the scalar filed potential and can locally create a bubble of the true vacuum, and thus trigger the decay of the false vacuum. For a critical bubble which is able to expand, such an interaction with the domain wall must be strong enough. 
\end{abstract}

%%%%%%%%%%%%%%%%%%%%%%%%%%%%%%%%%%%%%%%%%%%%%%%%%%

\maketitle
       
\section{introduction}

Domain walls are sheet-like topological  defects,  which may  be created in the early Universe when a discrete symmetry is spontaneously broken. A domain wall separates the space into two distinguishable vacuum states. The fundamental constants (like particle masses, fine structure constant etc) can change after crossing the domain wall \cite{Kibble:1976sj,Vilenkin:1982ks,Vilenkin:1984ib}. The energy density of the the domain wall network might dominate the total energy density of the Universe if the domain walls are generated at early times \cite{1975ZhETF..67....3Z}. This problem could be solved if the walls are unstable or using some other elaborate mechanism  \cite{Vilenkin:1981zs,Gelmini:1988sf,Larsson:1996sp,Dvali:1995cc,
Stojkovic:2005zh,Stojkovic:2004hz}. On the other hand, if the walls are formed in the late time epochs of the universe, they will be light and thus not easily detectable.  
Such domain walls can even play the role of dark matter candidates or generate other interesting structures \cite{Dai:2020rnc}.

Since domain wall can change the fundamental constants, there are many experiments invented to test their existence. These methods include mass difference causing acceleration \cite{McNally:2019lcg}, variance of the fine structure constant \cite{Roberts:2019sfo}, electric dipole moment measurements \cite{Stadnik:2014cea}, measuring magnetometers  \cite{Pospelov:2012mt,2013arXiv1303.5524P,JacksonKimball:2017qgk,Afach:2021pfd}, satellite synchronization  \cite{Derevianko:2013oaa,Roberts:2017hla,Kalaydzhyan:2017jtv} and 
gravitational wave detectors \cite{Hall:2016usm,Jaeckel:2020mqa,Grote:2019uvn}. 

The interaction between the domain wall and regular matter can be classified using a collection of gauge-invariant operators parameterizing  the Standard Model interactions with the fields that make up the composition of the wall (see e.g. \cite{Essig:2013lka}). However, in these studies matter was considered to be a fundamental point-like particle which will not distort the domain wall during the crossing. In contrast, this paper focuses on classical objects of finite sizes. Interaction of black holes with domain walls was studied in \cite{Stojkovic:2004wy,Frolov:2003mc,Frolov:2004wy,Frolov:2004bq,Frolov:1998td,Christensen:1998hg}, however no study was performed for finite but regular classical objects.  
 Earth or a satellite in Earth's orbit are in general such classical objects which, in principle, can distort a domain wall. We find that when a macroscopic object encounters the domain wall, it may be reflected or it may pass through depending on the strength of the interaction, relative velocity and size. During the encounter, the object loses its energy to deformations of the wall. In addition, the object changes its mass by passing from one side of the wall to another. 
We track motions of satellites in Earth's orbit with a great precision, so minor changes in satellites' parameters are observable in principle.  
For example, a typical velocity precision of a satellite is about $0.5$ mm/s, which directly puts an upper limit on its mass change to $\Delta M/M \lessapprox 5\times 10^{-17} $. Therefore, this effect can offer a new method for detection of domain walls by observing whether a satellite's velocity gets suddenly changed. The famous flyby anomaly -- anomalous increase in speed observed during a planetary flyby by a satellite --  is very similar to this effect. 

In the second part of the paper we show how matter can trigger the decay of the false vacuum. The presence of matter modifies the scalar filed potential and can locally create a bubble of the true vacuum. For a critical bubble which is able to expand, interaction with the domain wall must be strong enough.

\section{Interaction of a scalar field with classical objects}

Consider a solid massive spherically symmetric object, with a uniform mass distribution and radius $R$ in its own rest frame. The object is moving with velocity $v_O$ in $z$-direction. The location of the center of the object along the $z$-axis is labeled by $z_O$. The normalized mass distribution function of the object in axially symmetric (around $z$-axis) coordinates is         
\begin{equation} \label{dist}
f(z,r,z_O,v_O)=\Theta(R-\sqrt{r^2+\frac{(z-z_O)^2}{1-v_O^2}}) ,
\end{equation}       
 where $\Theta(x)$ is the Heaviside function. The radial coordinate is $r^2=x^2+y^2$, while $1/\sqrt{1-v_O^2}$ is the standard Lorentz contraction factor. The relativistic action for this object is 
\begin{equation}
S_m= -m\int \sqrt{1-v_O^2} dt ,
\end{equation}
where $m$ is the object's mass in the absence of any couplings. 

Consider now a domain wall made of a scalar field $\phi$. The scalar field's action is 
\begin{equation}
S_\phi =\int  \frac{(\partial_t \phi)^2}{2} - \frac{( \nabla \phi )^2}{2} - V(\phi ) d^4x  ,
\end{equation}
The coupling between the massive object and the scalar field can be described by 
\begin{equation}
S_{\rm int}= -A \int   f(\vec{r}) I(\phi) d^4x ,
\end{equation}       
where $A$ is the coupling constant, while $f(\vec{r})$ is given in Eq.~(\ref{dist}). $I(\phi)$ is a coupling function, which is model dependent. The equations of motion that follow from the total action $S_{\rm tot}= S_m+S_\phi+S_{\rm int}$ are 
\begin{eqnarray} \label{eom}
&&\partial_t^2 \phi-\partial_r^2 \phi -\frac{1}{r}\partial_r \phi-\partial_z^2 \phi+V'+A f(\vec{r})I'=0\\
&&\frac{d}{dt}  \Big(\frac{v_O}{\sqrt{1-v_O^2}} M\Big)=A\int_0^R (I(\phi_{d} )-I(\phi_{u})2 \pi rdr\\
&&M=m+A \int_0^R (I(\phi_{u})+I(\phi_{d}))\sqrt{R^2-r^2}2\pi rdr ,
\end{eqnarray}
where  $\phi_{u}=\phi(r,z_{u})$, $\phi_{d}=\phi(r,z_{d})$, with $z_{d}=z_O-\sqrt{R^2-r^2}\sqrt{1-v_O^2}$ and $z_{u}=z_O+\sqrt{R^2-r^2}\sqrt{1-v_O^2}$. The parameters $z_d$ and $z_u$ are the limits within which the object moves in $z$-direction. $M$ is the object's effective mass which is modified by the presence of $\phi$. The effective potential of the scalar field can be now written as 
\begin{equation}
V_{eff}=V+AfI .
\end{equation}

\subsection{Concrete model }

\begin{figure}
   %\centering
\includegraphics[width=8cm]{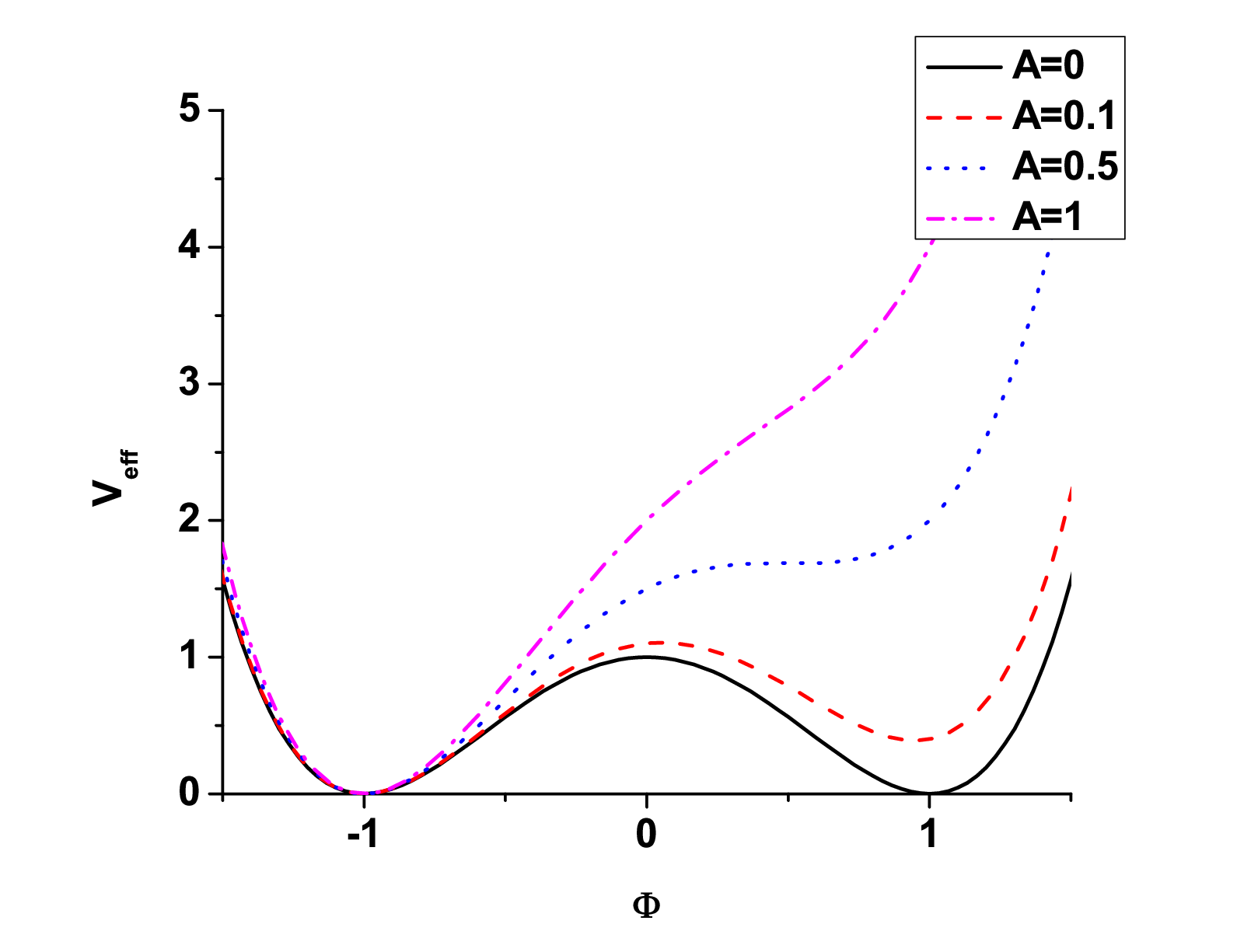}
\caption{ The effective potential of a scalar field which interacts with a massive object.  The solid line (no interaction) shows a potential with two equal minima at $\phi =\pm 1$. The potential and couplings are chosen as in Eqs.~\eqref{scalar-potential} and \eqref{I-term}  with $B=\Lambda =1$. The dashed line, $A=0.1$, shows a potential with the true vacuum at $\phi =-1$ and the false vacuum at $\phi=1$.  For larger values of $A$ there is no false vacuum.
}
\label{potential-mass}
\end{figure}

\begin{figure*}
\centering
\subfigure{
\includegraphics[width=0.4\textwidth]{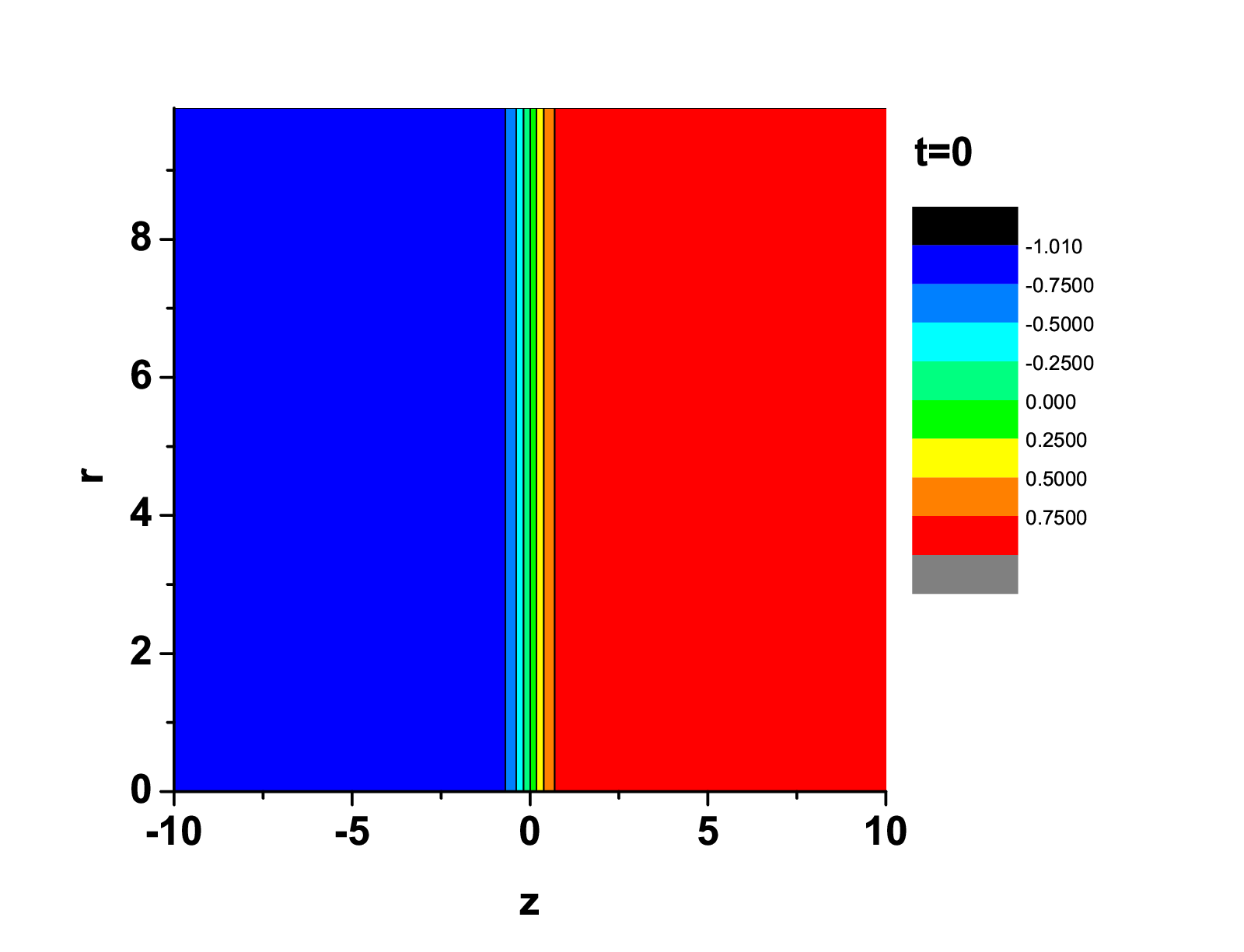} 
\includegraphics[width=0.4\textwidth]{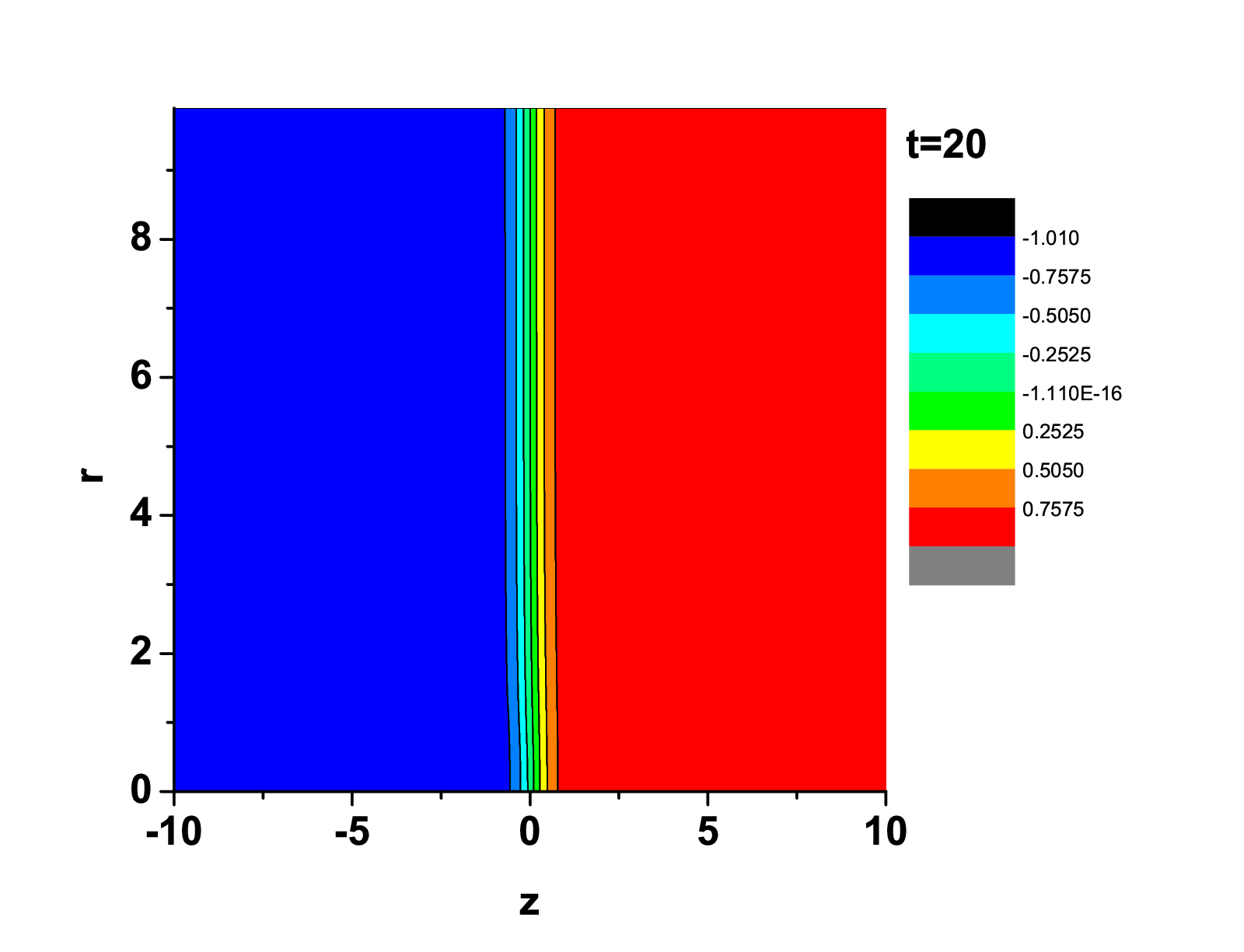} 
}
\subfigure{
\includegraphics[width=0.4\textwidth]{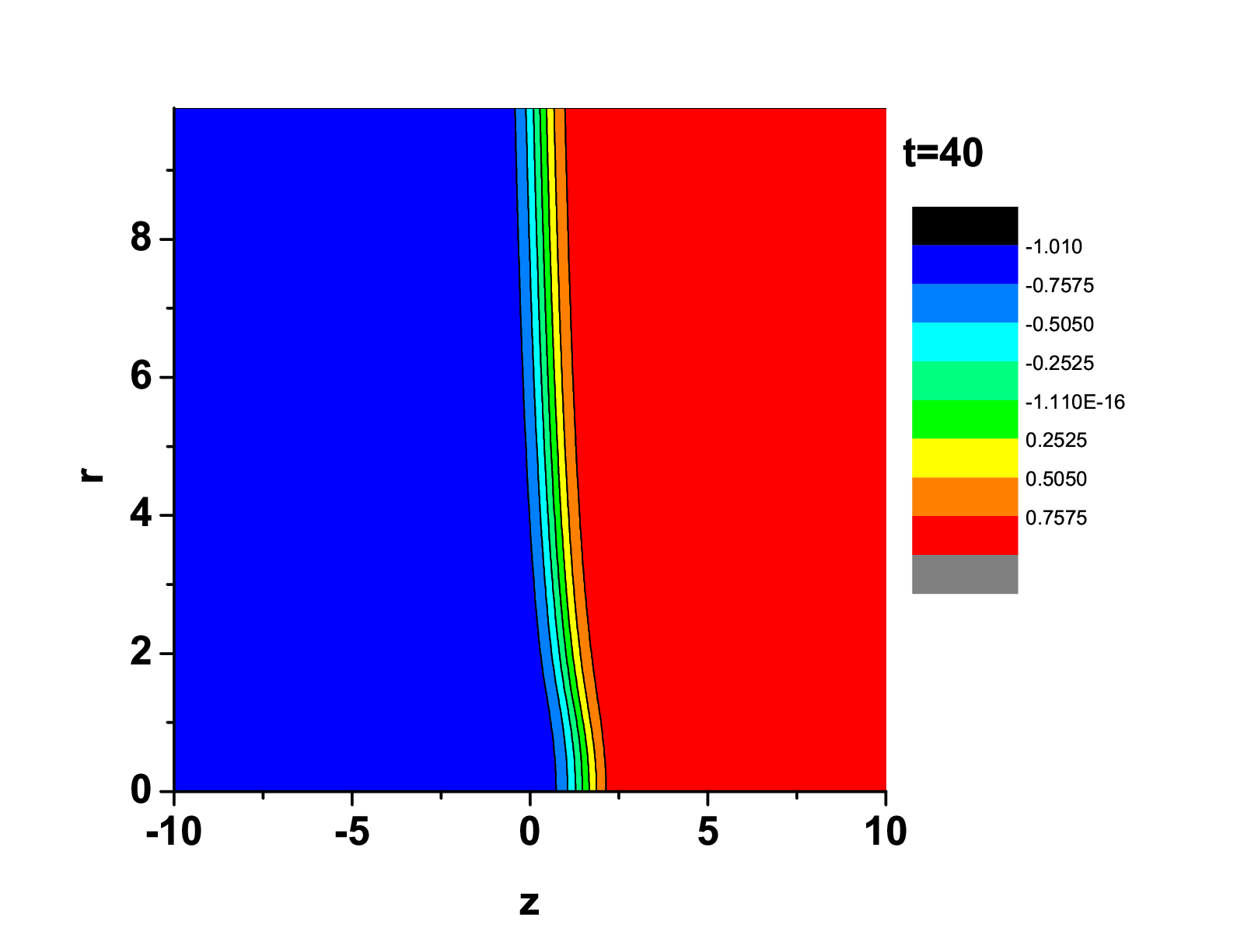} 
\includegraphics[width=0.4\textwidth]{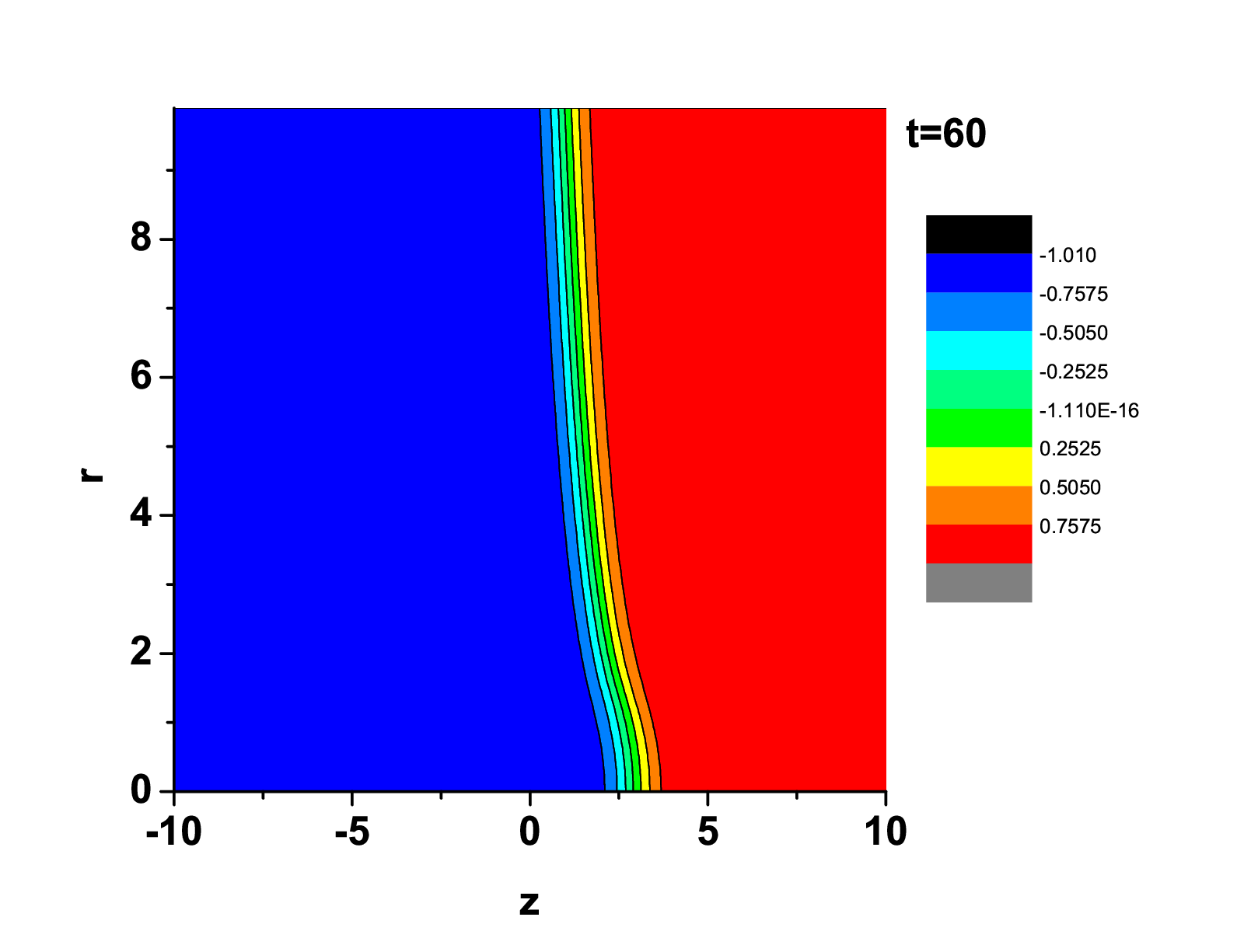} 
}
\subfigure{
\includegraphics[width=0.4\textwidth]{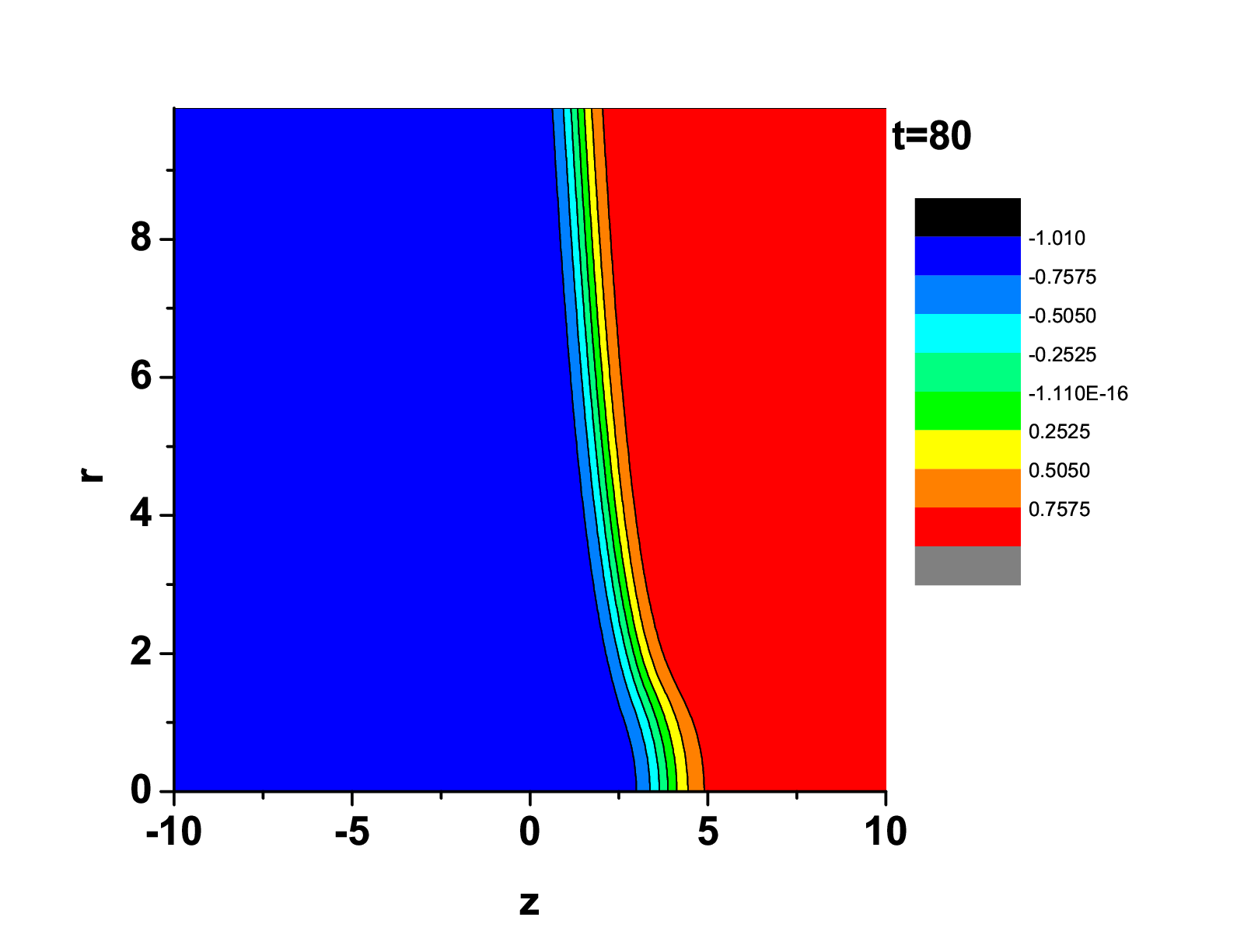} 
\includegraphics[width=0.4\textwidth]{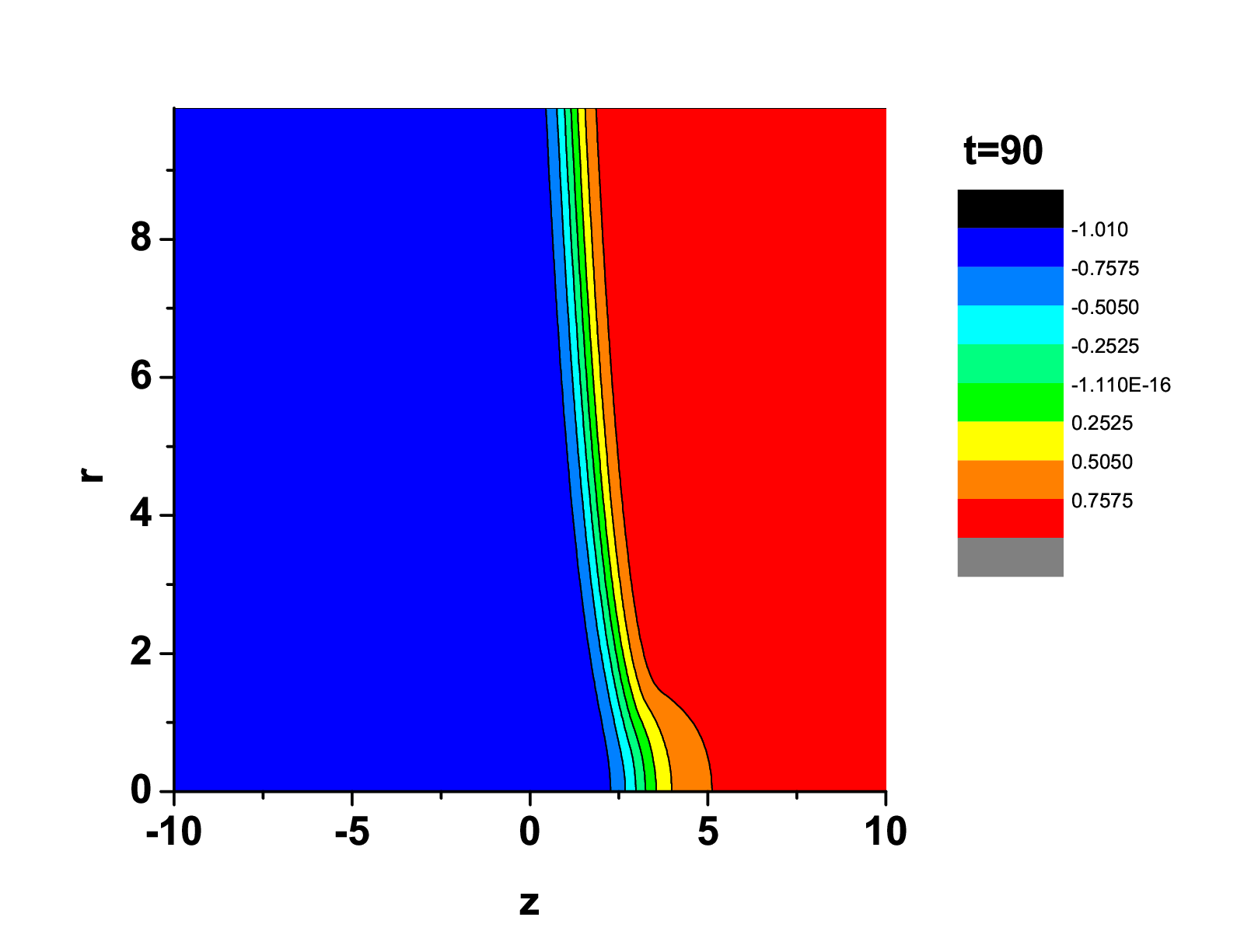} 
}
\subfigure{
\includegraphics[width=0.4\textwidth]{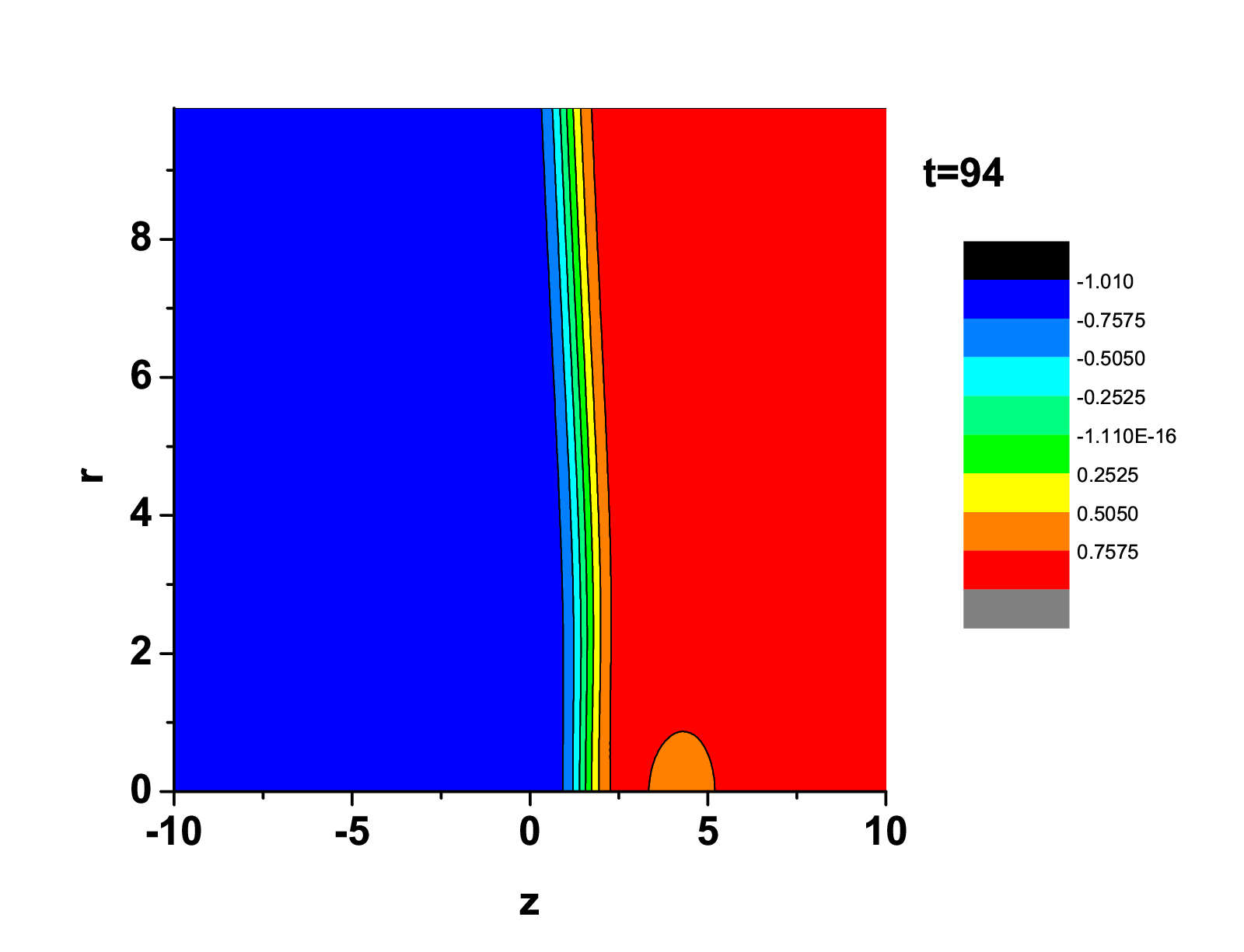} 
\includegraphics[width=0.4\textwidth]{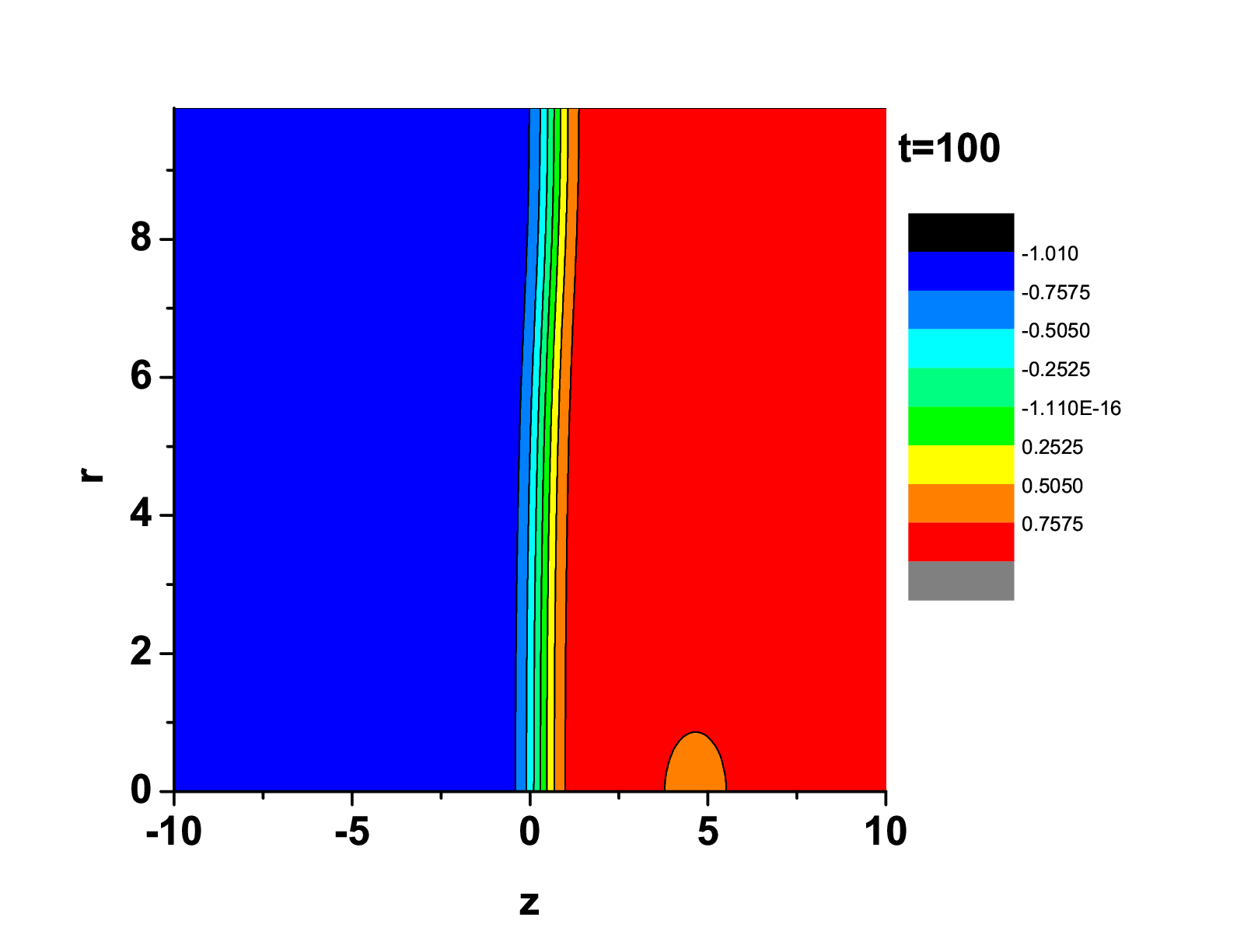} 
}
\caption{Distortion of the wall in cylindrical coordinates $(r,z)$, during the collision with a massive object moving in the $z$-direction. The color code represents the values of the scalar filed. The initial conditions are chosen to be $m=10^4$, $R=1.5$, $v_O=0.1$, $A= 0.5$ and $x_0 =-3.5$. At $t=0$, the domain wall is located at $z=0$ without any distortion. At $t=20$, the wall is pushed by the object to the right around $r=0$ and $z=0$. From $t =40$ to $t= 80$, the distortion becomes clearer. At $t=90$, the object starts to separate from the wall, where a significant bump can be observed. From $t=94$ to $t =100$, the object is completely separated from the wall, and a separate closed region with $\phi=\Lambda$ (new vacuum) is created. This is a  bubble of the scalar field formed around the massive object. The wall bounces back to its equilibrium configuration.  There are domain wall waves generated in this process, but they cannot be clearly seen in these plots because of the small amplitudes.  
}
\label{collide}
\end{figure*}

We choose to work with a well known scalar field potential that provides the existence of domain walls 
\begin{eqnarray}
\label{scalar-potential}
V= B (\phi^2-\Lambda^2)^2,
\end{eqnarray}
where $B$ and $\Lambda$ are constants. In particular, $\Lambda$ is the vacuum expectation value of the scalar field.
One dimensional kink and anti-kink solutions are described by the following profile
\begin{equation}
\label{wall-solution}
\phi=\Lambda \tanh (\pm \Lambda \sqrt{2B} \frac{z-vt-z_i}{\sqrt{1-v^2}}) ,
\end{equation}
where $z_i$ is the initial position of the kink, while $v$ is it's velocity. Without any loss of generality we boost the object to a frame where the domain wall is at rest initially, i.e. $v=0$, and its center is located at $z_i=0$. These are the initial conditions of the domain wall before it interacts with the massive object.  The kink solution is shown in the plot $t=0$ in Fig. \ref{collide}.  The energy per unit area of the domain wall is 
\begin{equation}
\rho=\frac{4}{3}\sqrt{2B}\Lambda^3.
\end{equation} 
We take the coupling function to be 
\begin{equation}
\label{I-term}
I=(\phi+\Lambda)^2 .
\end{equation}
With this choice, the object will not acquire any extra mass at $\phi=-\Lambda$ vacuum, so $M= m$ before the object encounters the wall. 
Fig.~\ref{potential-mass} shows that the presence of the object inevitably modifies the potential. The true and false vacuum states in empty space are different from the same states inside the matter distribution. The presence of matter can therefore change the scalar field's vacuum state.

\section{Method}

The equations of motion in Eq.~(\ref{eom}) must be solved numerically. We replace the derivatives with 
\begin{eqnarray}
\partial_t^2\phi&=&\frac{\phi (t+\Delta t)-2\phi(t)+\phi(t-\Delta t)}{\Delta t^2}\\
\partial_z^2\phi&=&\frac{\phi (z+\Delta z)-2\phi(z)+\phi(z-\Delta z)}{\Delta z^2}\\
\partial_r^2\phi&=&\frac{\phi (r+\Delta r)-2\phi(r)+\phi(r-\Delta r)}{\Delta r^2}\\
\partial_r\phi&=&\frac{\phi (r+\Delta r)-\phi(r-\Delta r)}{2\Delta r} ,
\end{eqnarray}
where $\phi(t=0)$, $\phi (t=\Delta t)$, $v_O(t=0)$ and $x_0(t=0)$ are given as initial conditions. The evolution of the field and object is obtained by iterations.   
We choose $\Delta r=\Delta z= 3\Delta t=10^{-2}$ to preserve the stability of the numerical method. The numerical range is $0\leq r\leq 30$ and $-30\leq z\leq 30$. The singularity at $r=0$ is avoided by replacing 
\begin{equation}
\lim_{r\rightarrow 0}\frac{1}{r}\partial_r \phi  =\lim_{r\rightarrow 0}\partial_r^2 \phi .
\end{equation} 
The integral term in Eq.~(\ref{eom}) is calculated using the Simpson method.

\section{Results}

We discuss two distinct cases. The first case is an object crossing the domain wall from one vacuum to another. The second case is an object triggering the vacuum decay by generating a bubble of the local vacuum state.

\subsection{Collision between a domain wall and a classical object}

\begin{figure}
   %\centering
\includegraphics[width=8cm]{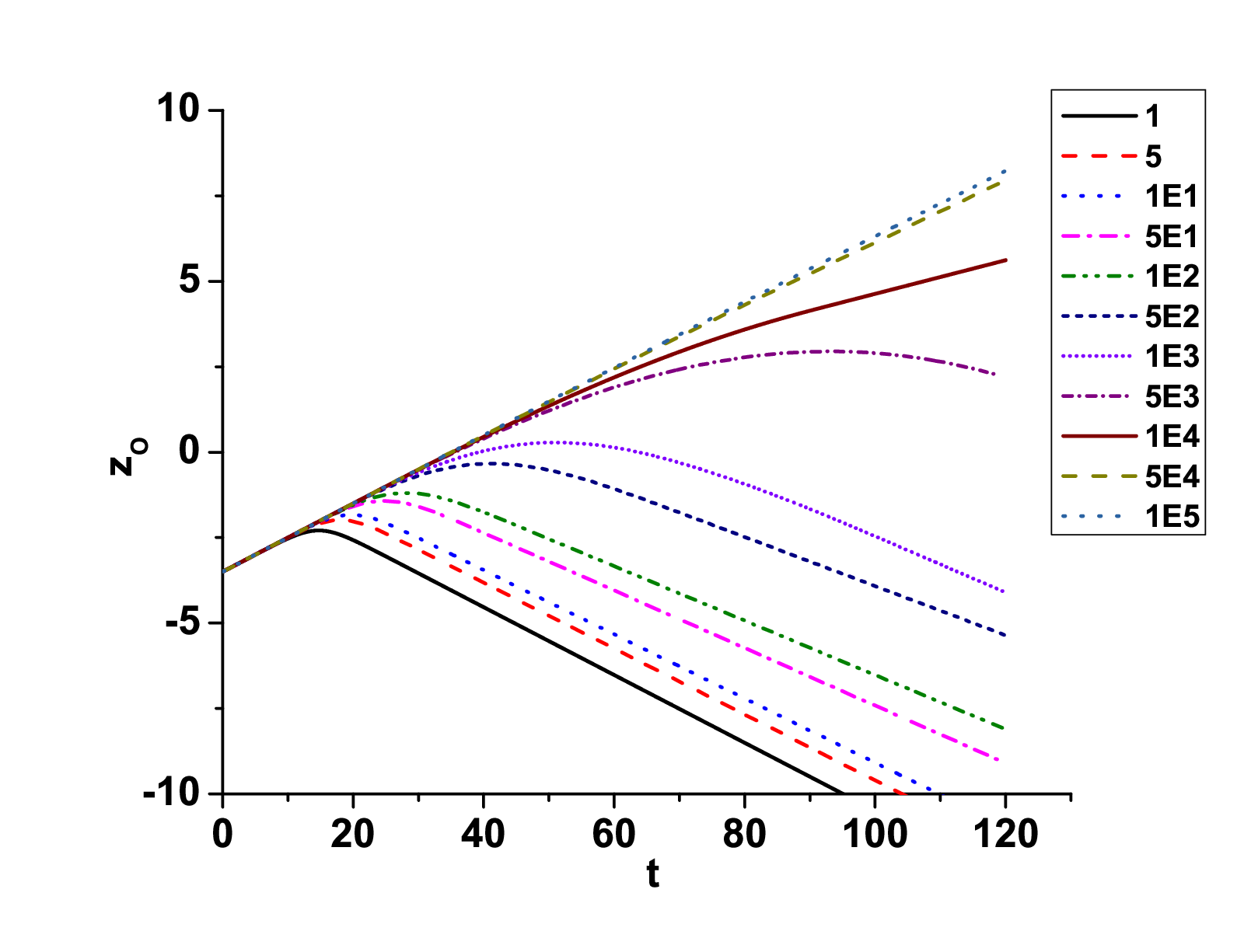}
\caption{ The path of the object during its interaction with the wall. On the y-axis we have $z_O$, which represents the location of the center of the object at some moment in time. The parameters are chosen to be $A=0.5$, initial $v_O=0.1$, $R=1.5$, while the values for the object's mass $m$ are marked in the plot. If the object is light, its momentum and energy are low, so it bounces back from the wall. As it becomes more massive, the wall bends more and more, up until the point where energy and momentum are high enough so that the wall cannot stop it any more, and the object passes through the wall.   
}
\label{path}
\end{figure}

We consider the scalar potential in Eq.~\eqref{scalar-potential} with $B=1$ and $\Lambda=1$.  The scalar field generates a domain wall structure  as in Eq.~\eqref{wall-solution}. In the absence of the interaction, i.e. $A=0$, Fig.~\ref{potential-mass} shows that the potential has two equal minima, which are at $\phi =\pm \Lambda$.
Initially, the domain wall is at rest, located at $z_i=0$. The $t=0$ plot in Fig.~\ref{collide} shows the scalar field with the domain wall structure. There is an obvious  transition at $z=0$. The vacuum state at $z<0$ ($z>0$) is $\phi=-\Lambda$ ($\phi=\Lambda$). Initially the object is at $z_O=-3.5$ and moves to the right with velocity $v_O$. The crossing starts when the object reaches the wall. The wall is pushed to the right and it is distorted (segments from $t=0$ to $t=90$ in Fig.~\ref{collide}). The wall is then pushed farther, and if the object has enough energy it crosses the wall, while the wall bounces back to its original place (segments from $t=90$ to $t=100$ in Fig.~\ref{collide}). If the object does not have enough energy, the object is reflected back (as described with $m<5000 $ case in Fig.~\ref{path}). In this process, the object changes its mass and velocity and its energy is applied to distorting the wall and generating waves, which cannot be seen directly in the plots in Fig.~\ref{collide}, because of their small amplitude. These waves can be seen in Fig.~\ref{true-vacuum} and \ref{false-vacuum}.   We note that we can chose the values for $R$ and $m$ independently since gravity is not included in the action. In other words, an object with $R=1$ and  $m=10^5$ is not necessarily within its own Schwarzschild radius.

\begin{figure}
   %\centering
\includegraphics[width=8cm]{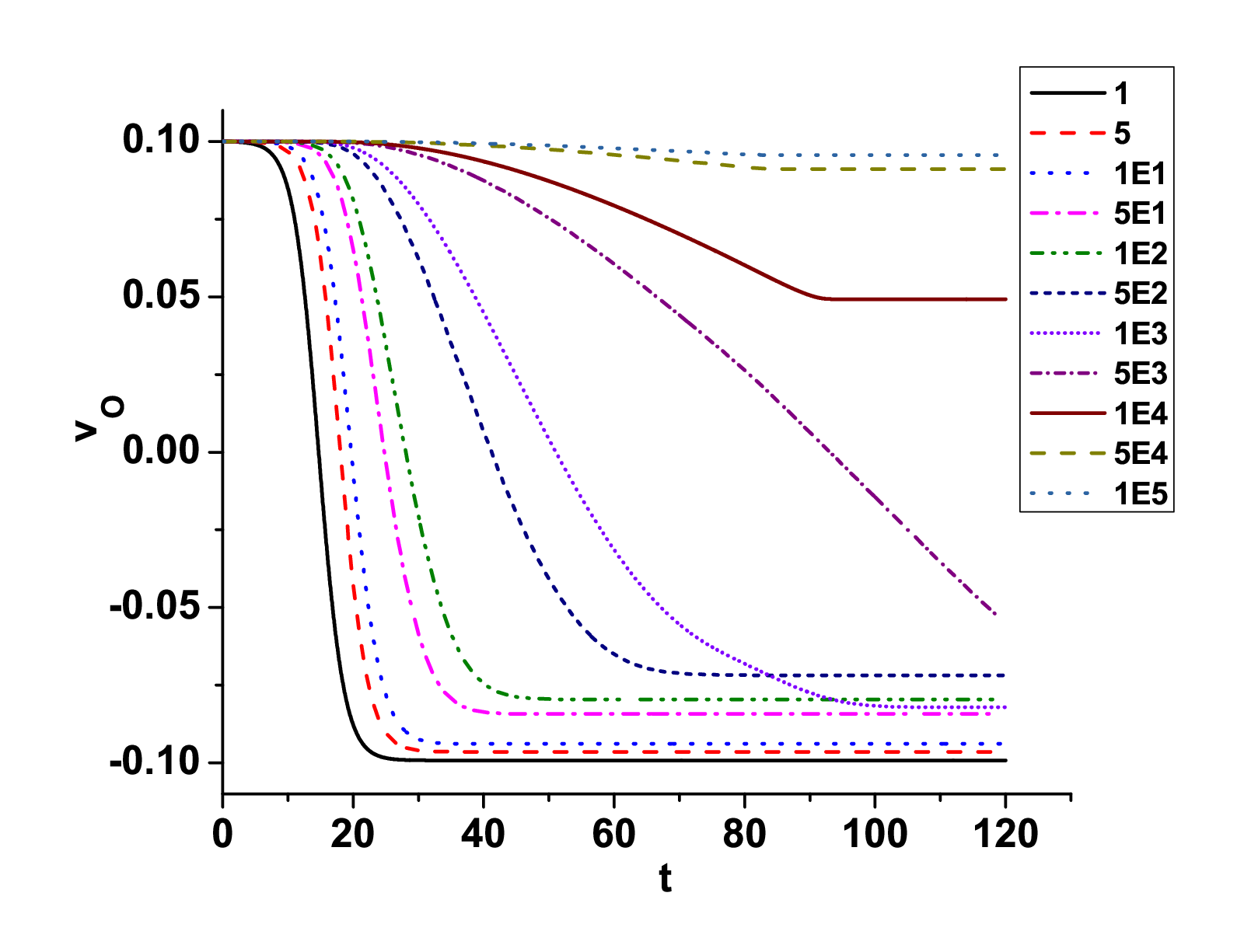}
\caption{ Evolution of the velocity of the object, $v_O$, during the collision with the wall as a function of time. The parameters are chosen to be $A=0.5$, initial $v_O=0.1$, $R=1.5$, while the values for the object's mass $m$ are marked in the plot. A very light object is bounced back (velocity is reversed) without any energy loss nor gain. Medium mass objects are reflected by the wall, but they also gain some wall distortion energy back during the collision. In these cases the wall acts like a spring. Therefore their velocity can be larger than in some of the lighter cases. Very massive objects pass through  the wall and cannot gain the distortion energy back, so they lose the highest amount of energy.     
}
\label{velocity}
\end{figure}

Fig.~\ref{velocity} shows how the velocity of an object changes in the process. Very light objects are reflected completely, which resembles the classical collision between a very light and very heavy object. As the object becomes more massive, it  distorts the wall more and more. The object's energy is applied toward the wall's distortion and wave dispersion. The reflected object's speed is lower than the initial speed. An object with a mass in a medium range may gain some distortion energy back when the wall is bounced back to its balanced position (e.g. $m=10^3$ case). In this case the object loses less energy than in some of the lower mass cases, since the wall acts like a spring. In contrast, very massive objects  just pass through the wall and do not recover any energy back from the distorted wall, so they lose more energy than less massive objects. 

\begin{figure}
   %\centering
\includegraphics[width=8cm]{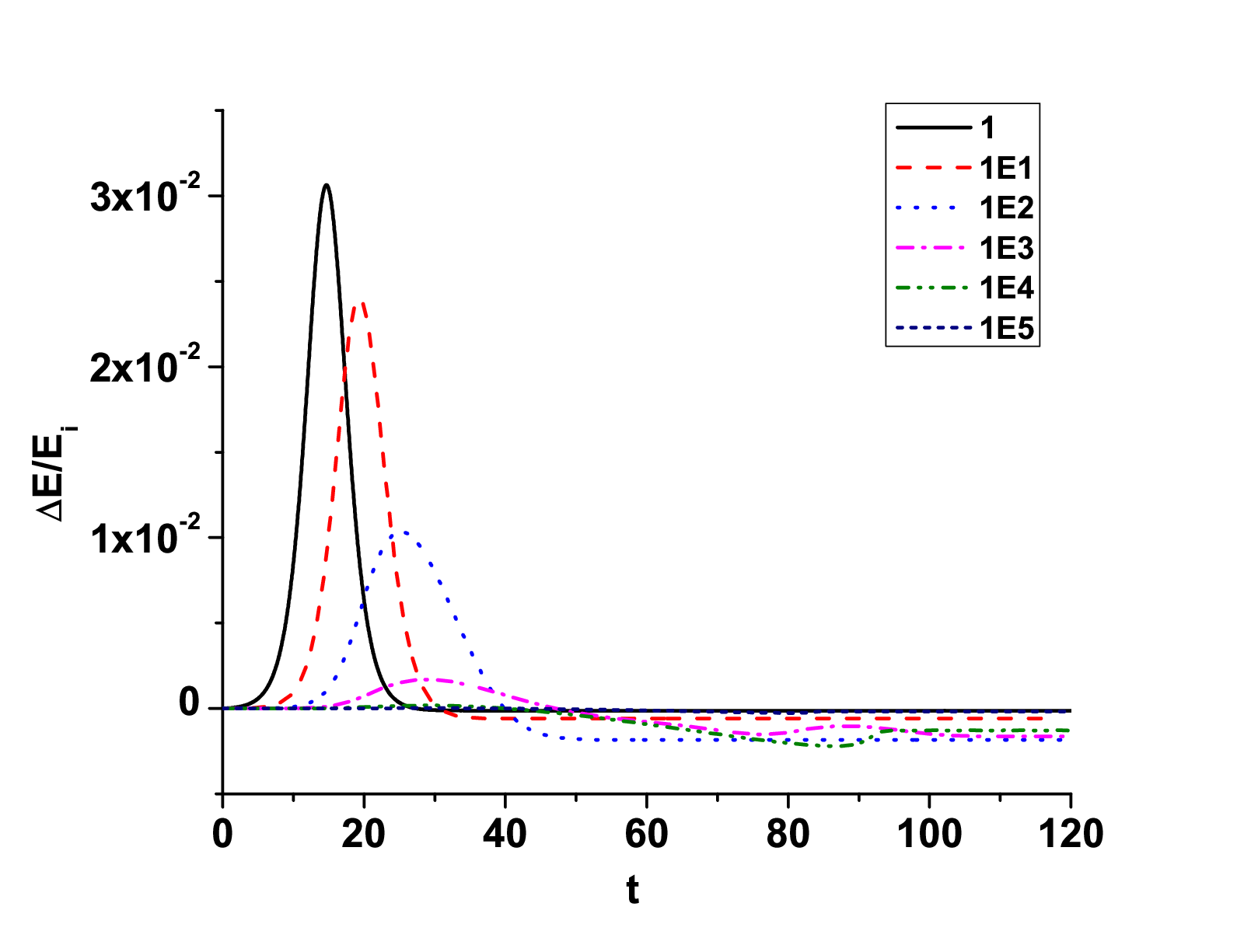}
\caption{ The relative change in energy of the object, $\Delta E/E_i$, in the collision process with the wall as a function of time. The parameters are chosen to be $A=0.5$, initial $v_O=0.1$, $R=1.5$, while the values for the object's mass $m$ are marked in the plot. As an object is approaching the wall its energy is increasing. Since the value of the scalar field at the object's location is increasing during the approach, the effective mass of the object, $M$, is increasing. Thus the energy of the object is increasing. During and after the interaction, the energy is decreasing since the velocity is decreasing, and also some of the energy is used to deform the wall. For medium mass cases, the object regains some energy when the wall bounces back (e.g. for $M=10^3$), so energy is increasing again at late times.       
}
\label{energy}
\end{figure}

Fig.~\ref{energy} shows how the energy of an object changes during the collision. The energy is calculated according to 
\begin{equation}
E_O=M/\sqrt{1-v_O^2},
\end{equation}
where $v_O$ is the velocity of the object.
At the beginning, the object gains some energy, because the value of $\phi$ is increasing and the effective mass, $M$, is increasing along with it. Then this gain is not sufficient to compensate for the loss, and more and more energy is released to the wall distortion and waves. At the end, the lost energy is higher than the energy gained from the wall. This lost energy does not come back to the object.

\begin{figure}
   %\centering
\includegraphics[width=8cm]{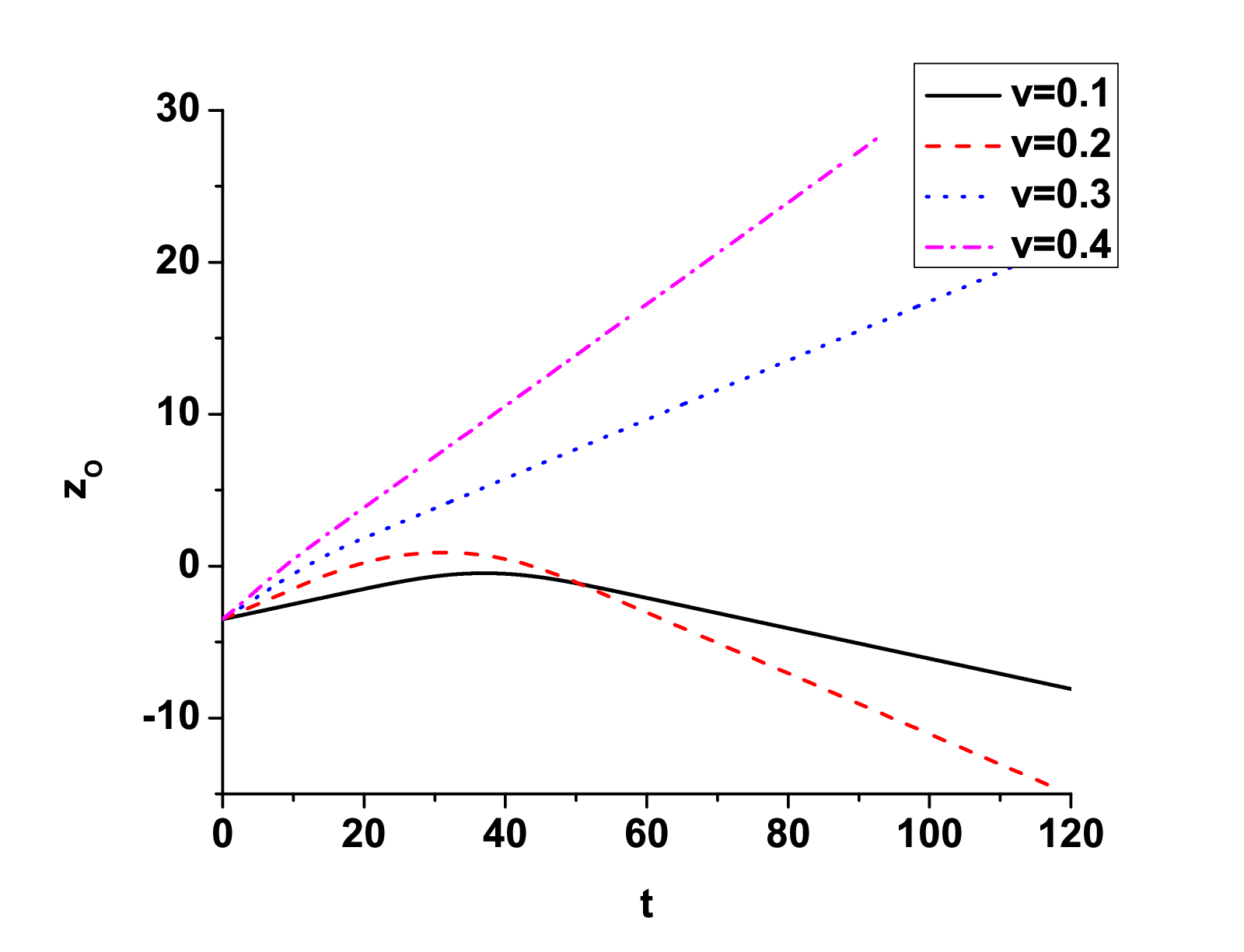}
\caption{The path of the object in the collision process with the wall as a function of its velocity. On the y-axis we have $z_O$, which represents the location of the center of the object at some moment in time. The parameters are chosen to be $A=5\times 10^{-4}$, $m=1$, $R=1.5$, while the initial values for initial $v_O$ are marked in the plot. For low velocities the object cannot cross the wall. For high enough velocities the object has enough energy to pass through the wall.   
}
\label{vary-velocity}
\end{figure}

The velocity of the object plays twofold role during the crossing. First, the velocity increases available kinetic energy, and second, it causes the Lorentz contraction which also changes the interaction between the object and wall. Fig.~\ref{vary-velocity} shows that only objects with high enough velocity can cross the wall. 

\begin{figure}
   %\centering
\includegraphics[width=8cm]{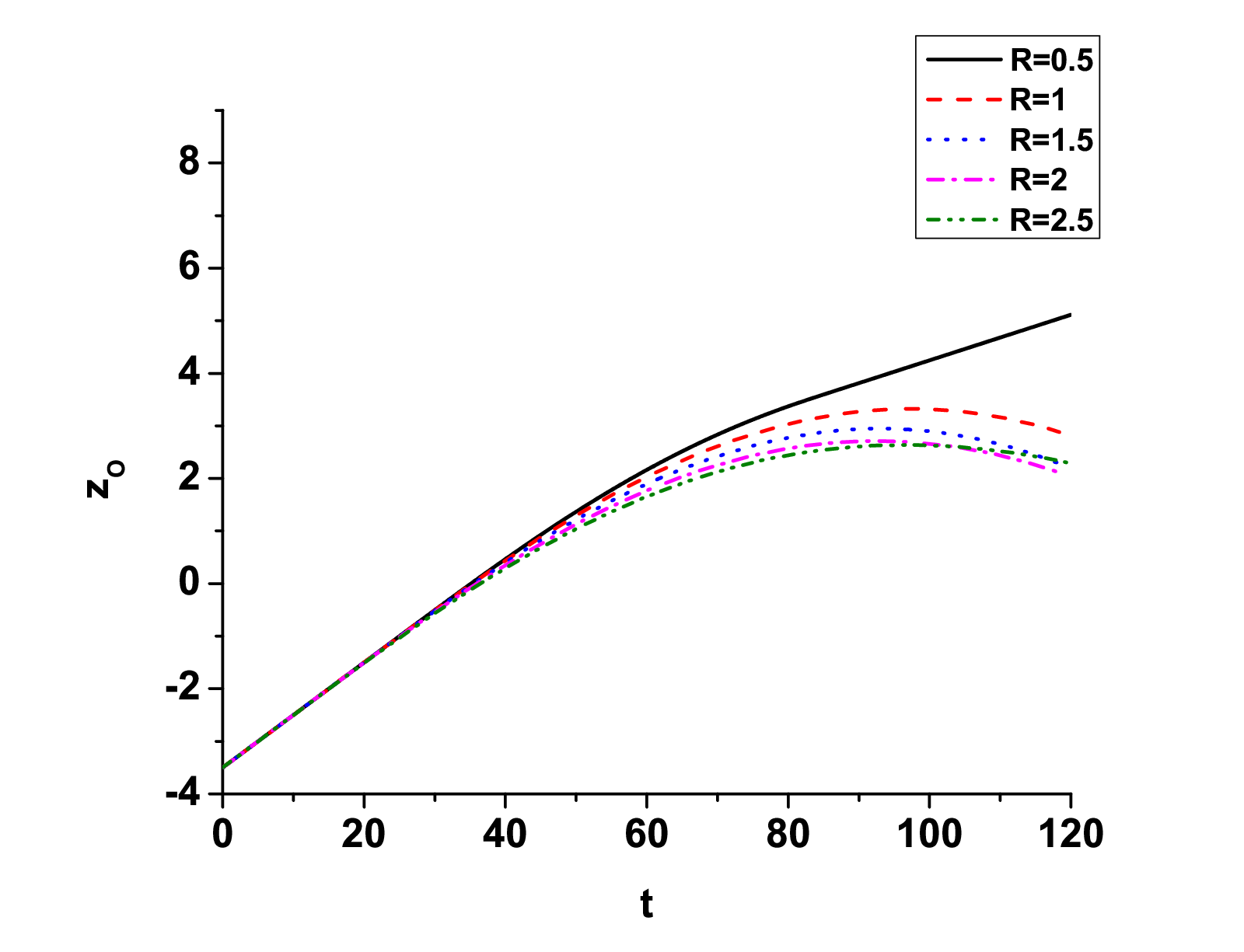}
\caption{The path of the object in the collision process with the wall as a function of its radius. On the y-axis we have $z_O$, which represents the location of the center of the object at some moment in time. The parameters are chosen to be $m=5000$. $A=0.5$. initial $v_O =0.1$, while the values for $R$ are marked in the plot. If the radius of the object is small the resistance of the wall is small, and it is easier to go through the wall. Objects of larger radii create greater distortion of the wall and lose more energy.  
}
\label{radius}
\end{figure}

Fig.~\ref{radius} shows how the motion of the object varies as a function of its radius. If the radius is small, the wall resistance is smaller, and the object easier penetrates the wall. Objects of larger radius induce greater distortion during the crossing, and in turn more energy is lost.

\subsection{Vacuum decay in the presence of a massive object} 
\begin{figure}
   %\centering
\includegraphics[width=8cm]{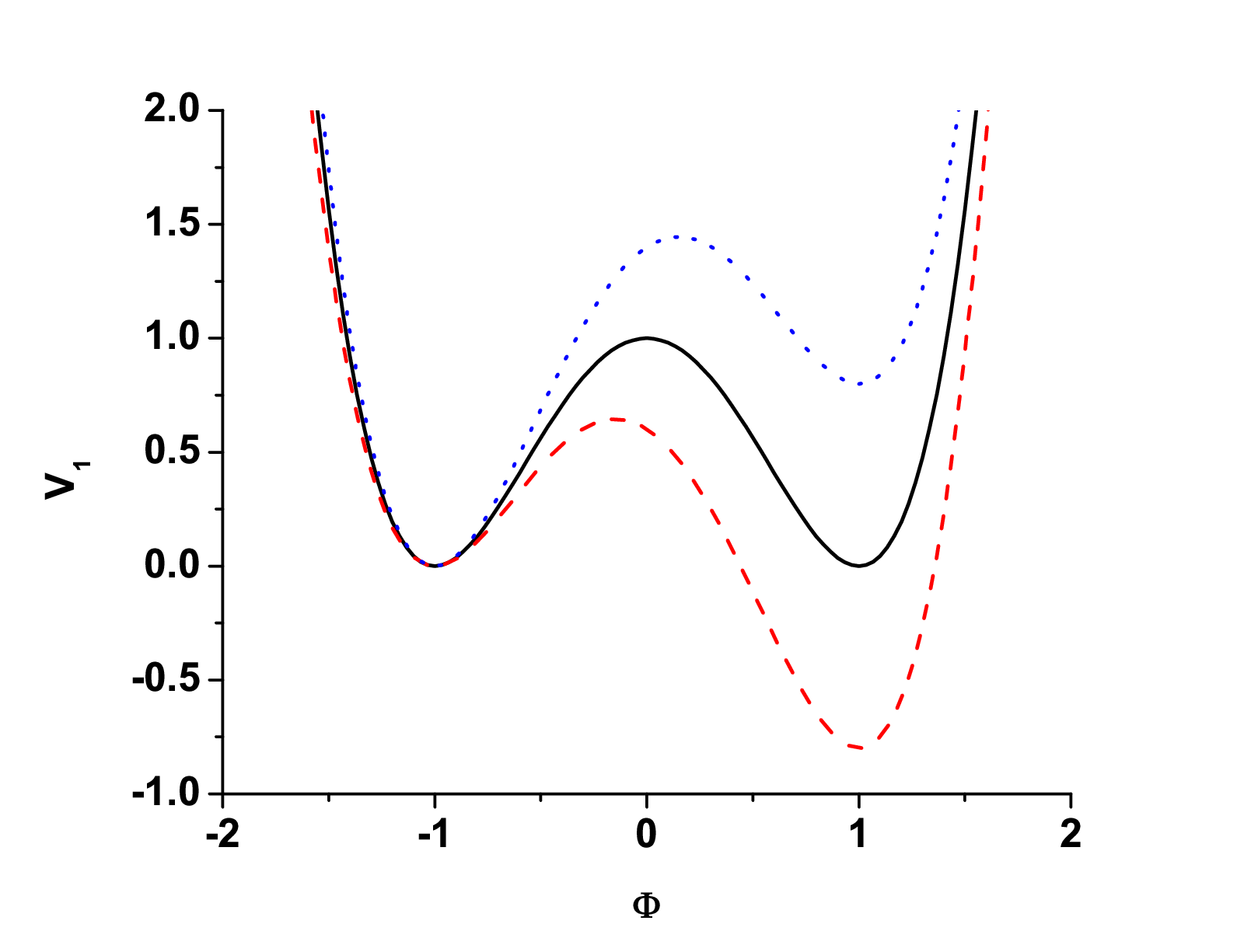}
\caption{The potential of a scalar field in the presence of a massive object as given in Eq.~\eqref{VTF}. We set $\Lambda=B=1$. The solid line, $a=0$, shows the potential with two equal minima at $\phi =\pm 1$. The doted line, $a=-0.2$, shows the potential with a false vacuum at $\phi =1$ and a true vacuum at $\phi=-1$.  The dashed line, $a=0.2$, shows a potential with a false vacuum at $\phi =-1$ and a true vacuum at $\phi=1$. 
}
\label{potential-decay}
\end{figure}

The presence of matter can change the scalar field effective potential, as we have seen in Fig.~\ref{potential-mass}. The true vacuum can be modified by the matter distribution and become a false vacuum locally. The opposite is also true, a region of false vacuum can be converted into a true vacuum state. In other words, the presence of matter can trigger a vacuum decay. To study this effect, we add an extra term in the scalar filed potential which breaks the $\phi \rightarrow -\phi$ symmetry and introduces the difference in the vacua 
\begin{equation}
\label{VTF}
 V=B(\phi^2-\Lambda^2)^2+ a (\phi^3-3\Lambda^2 \phi -2\Lambda^3) .
\end{equation}
This potential is plotted in Fig.~\ref{potential-decay}. We see that for $a=-0.2$ ($a=0.2$), $\phi=\Lambda$($\phi=-\Lambda$) is the false vacuum expectation value, while  $\phi=-\Lambda$($\phi=\Lambda$) is the true vacuum expectation value. 

\begin{figure}
   %\centering
\includegraphics[width=8cm]{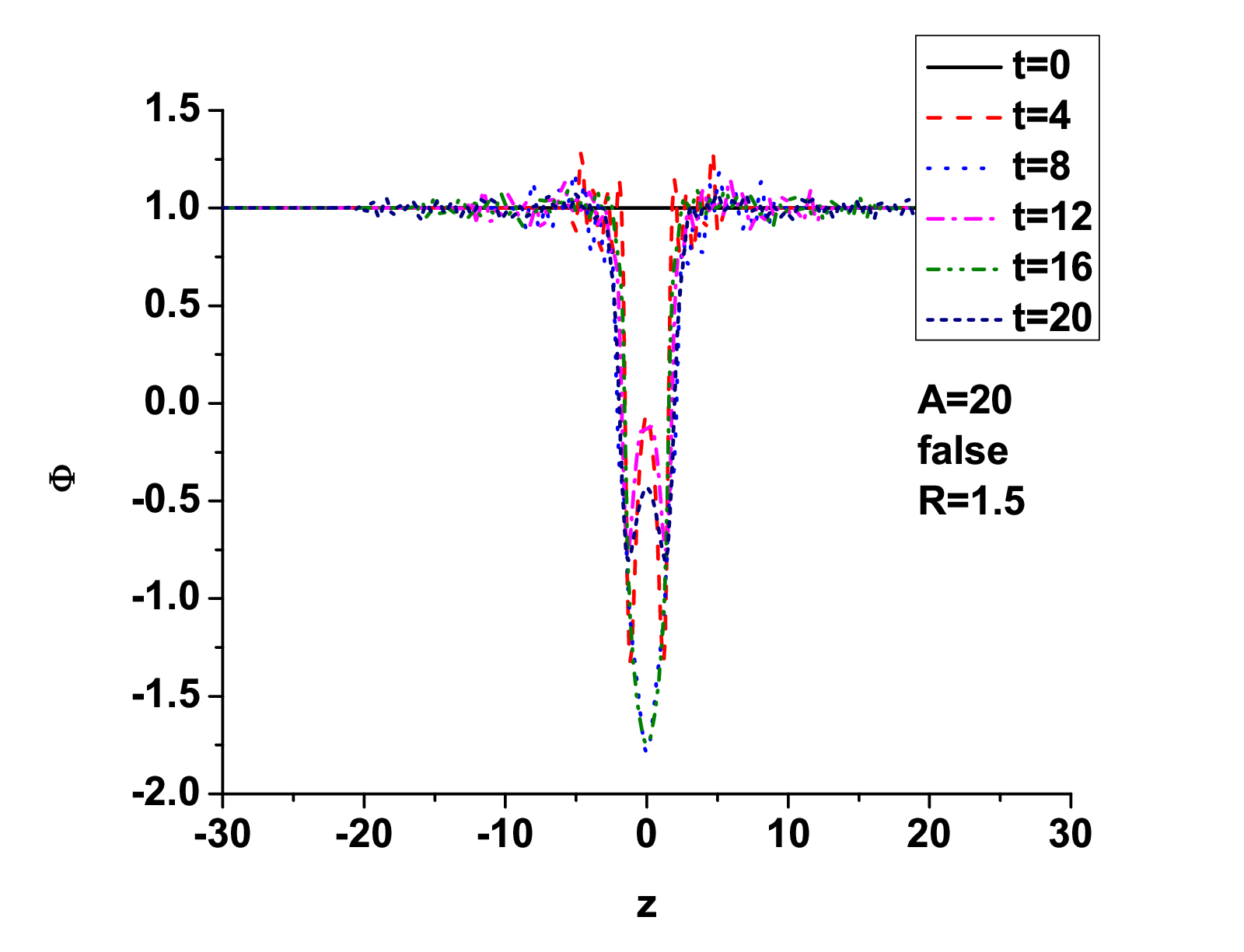}
\caption{ A massive object affecting the vacuum in its neighborhood. We set initial $v_O=0$,$A=20$, $R=1.5$, $\Lambda =1$, $B=1$, $m= 10^{10}$, $a=0.2$, $r=0$. The whole space was initially occupied by $\phi=\Lambda$. The scalar field inside the object evolves toward the lower values, because in this region the true vacuum is no longer at $\phi=\Lambda$. Outside of the object the field stays in the true vacuum at $\phi=\Lambda$. In this process, the waves are generated which propagate away and take energy away from the location.
The scalar field potential is modified only locally (in the bubble) around the object.  
}
\label{true-vacuum}
\end{figure}

Consider the region which is in the true vacuum ($a=0.2$ and $\phi=\Lambda$). The presence of the massive object will perturb the space and modifies the scalar field potential. Fig.~\ref{true-vacuum} shows that the scalar field is in the $\phi=\Lambda$ state at the beginning. The scalar field inside the object evolves toward the lower values, because in this region the true vacuum is no longer at $\phi=\Lambda$. Outside of the object the field stays in the true vacuum at $\phi=\Lambda$. At the same time the waves are generated which propagate away and take energy away from the location. The object mostly changes the vacuum state in its own neighborhood (within the bubble which is formed around it), while the energy from the vacuum decay is released. This is a clear demonstration that a massive object can affect the vacuum state around it. 

\begin{figure}
   %\centering
\includegraphics[width=8cm]{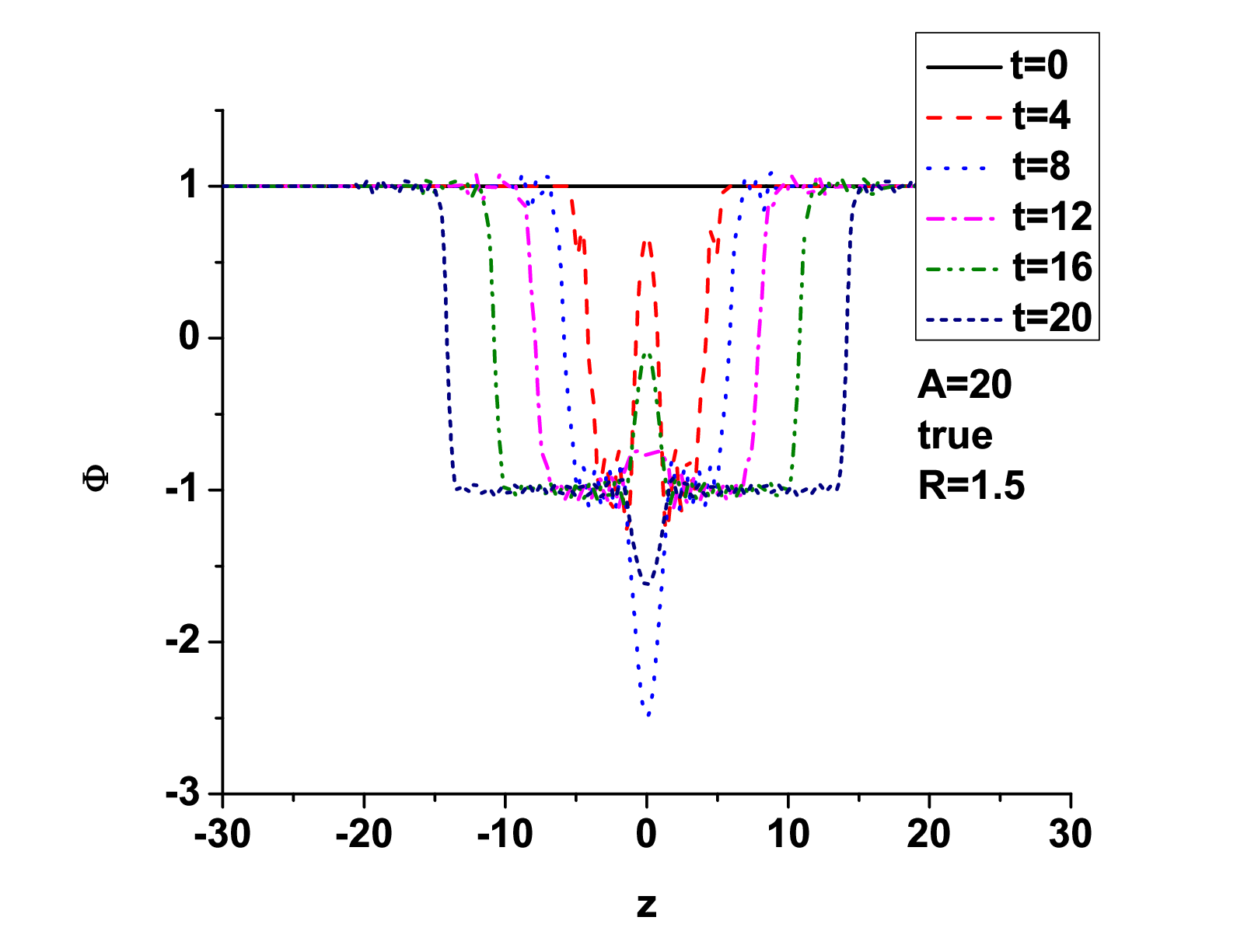}
\caption{ A massive object triggering a false vacuum decay. We set initial $v_O=0$, $A=20$, $R=1.5$, $\Lambda =1$, $B=1$,  $m= 10^{10}$, $a=-0.2$, $r=0$. The whole space was initially occupied by $\phi=\Lambda$. The scalar field starts to evolve toward the lower values (true minimum)  inside the object. The crucial difference from the previous case in Fig.~\ref{true-vacuum} is that after the transition starts, the region converted into the true vacuum grows. In other words, the true vacuum bubble expands, which means that a massive object can trigger (or catalyze) the vacuum decay.
}
\label{false-vacuum}
\end{figure}

\begin{figure}
   %\centering
\includegraphics[width=8cm]{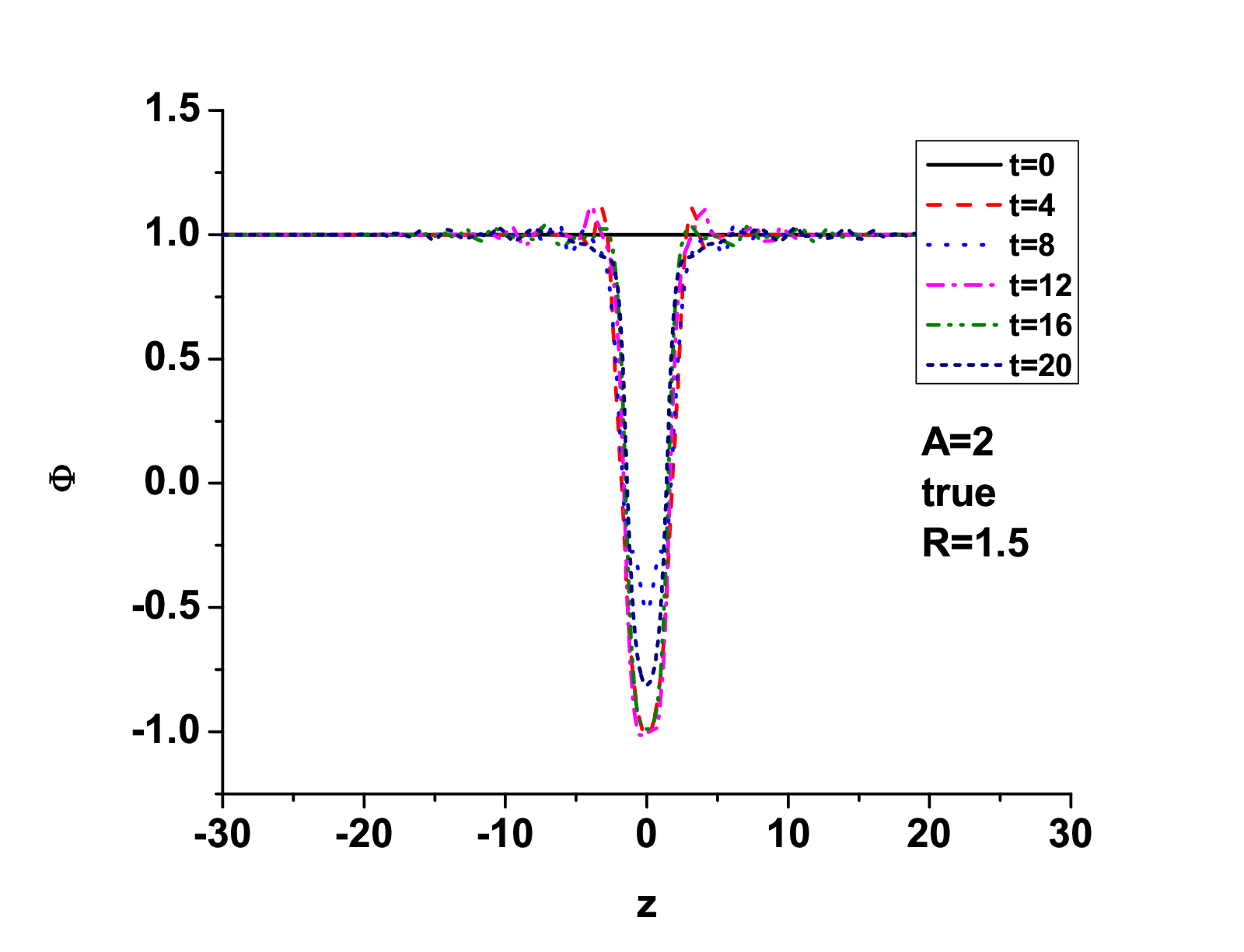}
\caption{Not all of the bubbles of the true vacuum are able to expand. If there is no enough energy to support the expansion of  the bubble (e.g. if the strength of the interaction between the object and the field is low), the vacuum decay process cannot be completed. We set here initial $v_O=0$, $A=2$, $R=1.5$, $m= 10^{10}$, $a=-0.2$, $r=0$.  Note that the value of the parameter $A$ is much lower than in Fig.~\ref{false-vacuum}.
}
\label{decay-not}
\end{figure}

Consider now the opposite situation, i.e. the region of space which is in the false vacuum ($a=-0.2$ and $\phi=\Lambda$) at the beginning. The presence of the massive object again perturbs the space and modifies the scalar field potential. Fig.~\ref{false-vacuum} shows that the scalar field is in the $\phi=\Lambda$ state at the beginning, but it starts to evolve toward the lower values inside the object. The crucial difference from the previous case is that after the transition starts, the region converted into the true vacuum grows. In other words, the true vacuum bubble expands, which means that a massive object can trigger (or catalyze) the vacuum decay. At the end of the process the whole space could be converted into the true vacuum. Whether this will happen or not depends on whether there is enough energy to support the the domain wall bubble expansion. Fig.~\ref{decay-not} shows that if the strength of the interaction (determined by the parameter $A$) is low, then the true vacuum will be created only locally around the object. This happens because the scalar field still sits in the region where the potential is higher than the false vacuum outside the matter, though inside the matter the scalar field may reach its true minimum. In that case, the bubble needs some extra energy source to keep expanding. 

\begin{figure}
   %\centering
\includegraphics[width=8cm]{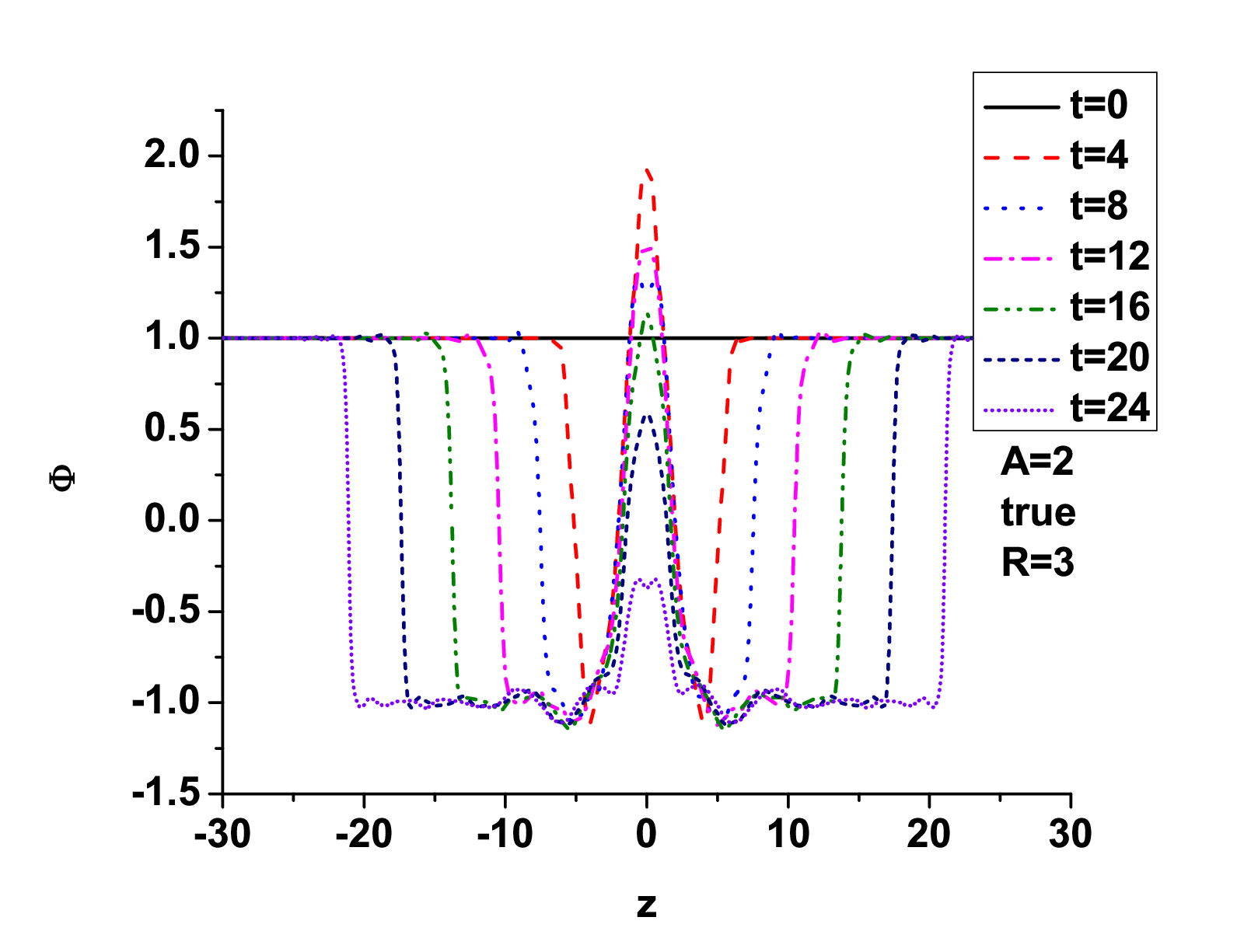}
\caption{ If the massive object is large enough, the bubble of true vacuum will expand. We set here initial $v_O=0$, $A=2$, $R=3$, $m= 10^{10}$, $a=-0.2$, $r=0$. 
}
\label{decay-large}
\end{figure}

In other words,  an expanding domain wall bubble must generate more energy than that released from the difference in vacua, or the scalar field cannot freely settle down into the true vacuum in the region outside of the matter. The condition for the false vacuum decay to release enough energy to support an expanding bubble is  
\begin{equation} 
4\pi R_b^2 \frac{\Delta \phi^2}{L}\lessapprox \frac{4\pi}{3} R^3\Delta V  ,
\end{equation}
where $L$ is the domain wall width, $R_b$ is the domain wall bubble radius, while $\Delta V$ is the potential difference between the false and true vacuum. The left side represents the surface energy of the domain wall bubble, while the right side is the volume energy released by the vacuum decay.  
If the volume of the matter distribution is large enough, then a large enough region populated by the scalar field will be pushed to its true vacuum, and once it passes the potential barrier outside the matter distribution, the false vacuum decay will become spontaneous. Figs.~\ref{decay-not} and \ref{decay-large} demonstrate how larger massive objects can trigger a successful vacuum decay easier than smaller ones. Higher values of the coupling parameter, $A$, can also trigger vacuum decay easier (as shown in Figs.~\ref{false-vacuum} and \ref{decay-not}), because more energy is generated in the process.

\section{Conclusions}

On the quantum level, a microscopic particle can either tunnel through or be reflected by a domain wall. However, a large classical object cannot tunnel through a domain wall, so it can either pass through, if it has enough energy, or bounce back, if it is not energetic enough. Obviously, a classical object encountering a domain wall must be treated differently from a microscopic particle. 

In our analysis, we considered a very general Lagrangian for a massive classical object interacting with a domain wall. By solving the equations of motion, we showed that the domain wall gets distorted as the classical object encounters it. The degree of distortion depends on the object's mass, velocity and size, and the strength of the interaction between the object and the wall. Some of the energy of the object is dissipated away by exciting waves on the wall. To cross the wall, the object must have enough energy to overcome the wall distortion and energy dissipation. Otherwise, the object will rebound and gain some energy back, since the wall will act as a spring in that case. 

Our results imply that if Earth (or any other planet), or a satellite in Earth's orbit, crosses the domain wall, it will lose (or under certain conditions gain) energy and momentum, and change its mass. One may then use this fact  to put a constraint on theories that allow for the existence of domain walls by studying orbits of planets or satellites.  

A sub-decimeter position accuracy and a sub mm/s velocity accuracy can be achieved in a ground-based reduced dynamic orbit determination using dual-frequency carrier phase measurements along with precise GPS ephemeris products and auxiliary environmental information. Here we adopt 0.5mm/s as the relevant 
velocity precision \cite{Montenbruck}. When the satellite mass changes when it passes through the wall, its velocity will change too. If the interaction is weak, we may estimate the mass difference from energy conservation
\begin{equation}
\frac{M_f}{1-v_f^2} \lessapprox \frac{M_i}{1-v_i^2}  ,
\end{equation}
where the subscripts $f$ and $i$ label the final and initial states respectively. Since the satellite's velocity is of the order of $10$km/s, the satellite's mass change is  
\begin{equation}
\frac{\delta M}{M}\lessapprox v\delta v \approx  5\times 10^{-17} .
\end{equation}
We note that this is an underestimate of the effect because the domain wall will also be distorted which will in turn also cause energy and momentum change of the satellite. 
Therefore, the existence of any domain wall that causes satellite mass distortion greater than this is already excluded.  
This method is different from using atomic clocks to constrain domain wall crossing \cite{Derevianko:2013oaa,Roberts:2017hla}. 

On the other hand, it is known that velocities of some satellites deviate from theoretical predictions when they pass their perigee. This is known as the flyby anomaly \cite{2017arXiv170402094S}. Such an effect can easily be explained by the satellite crossing a domain wall. In Fig.~\ref{collide} we saw that a closed domain wall (bubble) can be formed when a massive object like Earth passes through the domain wall.  A boundary of the bubble is also a domain wall of the same kind. Depending on the exact radius and shape of the bubble, a satellite's orbit might or might not intersect the bubble. This might explain why the anomaly does not show up consistently for all the satellites all the time.  Since the typical magnitude of the 
flyby anomaly is $\delta v \sim 10 mm/s$, the relative mass change of  $\frac{\delta M}{M} \sim 10^{-15}$ would be sufficient to explain the anomaly. 

The timescale on which the satellite changes its mass depends on the satellite crossing time, or more precisely, on a relative satellite-wall velocity and the thickness of the wall. The thickness of the wall is determined by the parameters in the potential, i.e. constants $B$ and $\Lambda$, however it is a priori unknown.  
Satellite data is typically collected in $30$ seconds intervals \cite{Montenbruck}. Since the orbital satellite velocity around earth is $\sim 10 $km/s, a relative satellite-wall velocity is perhaps anywhere from tens to hundreds of km/s. Thus, this method would work the best for domain walls less thick than $3000$ km. For thicker domain walls, the change would be slow and continious,  and some more sophisticated modelling would need to be performed.   
If the earth is surrounded by a spherical domain wall, then even a single satellite might be enough to detect the wall. The best chance would have a satellite with an elliptic orbit, since it would have a better chance of crossing the wall. Tracking motion of a network of satellites certainly increases the chance of detection. 
 On the other hand, the Pioneer data should also be very useful since they span several decades and very large distances. For this reason, the Pioneer mission is more convenient for detecting a network of domain walls than a single satellite.

We also showed than a massive object can trigger vacuum decay  by creating a bubble of true vacuum around it.  However, to create a critical bubble which is able to expand, enough energy must be released in this process. The released energy depends on the strength of the coupling between the massive object and the scalar field, and also on the size of the object. A similar effect was described in \cite{Burda:2015isa}  where black holes trigger the vacuum decay.

\begin{acknowledgments}
D.C Dai is supported by the National Natural Science Foundation of China  (Grant No. 11775140).
D. M. is supported in part by the US Department of Energy (under grant DE-SC0020262) and by the Julian Schwinger Foundation. D.S. is partially supported by the US National Science Foundation, under Grant No.  PHY-2014021. We thank Yu Sang for very useful suggestions. 
\end{acknowledgments}

\end{document}